\documentclass[12pt,prl,showkeys]{revtex4}
\usepackage{amsmath}
\usepackage{amssymb}
\usepackage{pst-node}
\usepackage{pstricks}
\newcommand{\be}{\begin{equation}}
\newcommand{\ee}{\end{equation}}
\newcommand{\ba}{\begin{eqnarray}}
\newcommand{\ea}{\end{eqnarray}}
\newcommand{\nn}{\nonumber}
\begin{document}

\preprint{hep-ph/0305331}

\title{Origin of quark masses and Cabbibo-Kobayashi-Maskawa matrix in a gauge theory with nonunitary parallel transporters
}

\author{C. Lehmann}
\email{claudia.lehmann@desy.de}

\author{G. Mack}
\email{gerhard.mack@desy.de}

\author{T. Pr\"ustel}
\email{thorsten.pruestel@desy.de}

\affiliation{ II. Institut f\"ur Theoretische Physik, Universit\"at Hamburg}
\homepage{http://lienhard.desy.de/}

\date{30 May 2003}



\begin{abstract}
Starting from a characteristic biinvariance property of Higgs potentials in gauge theories with nonunitary parallel transporters, we explain how quarks of different flavor can acquire different masses by spontaneous symmetry breaking and what is the difference between colour and flavor. We present a gauge theoretic model where the Cabbibo-Kobayashi-Maskawa matrix becomes computable in principle.
The model lives on a five dimensional space time with
  two four dimensional boundaries $R$ and $L$ a small distance $d$
  apart. Right handed quarks and leptons live on $R$ and left handed
  quarks and leptons on $L$. Photons can propagate in the bulk.
ee- and deep inelastic eN-scattering at energies $>d^{-1}$ will be affected.
\end{abstract}

\maketitle
Parallel transporters (PT) are of fundamental importance in gauge theory. 
In particular, lattice gauge fields are parallel transporters along links $<x,y>$ of a lattice; they are unitary maps between vector spaces at $x$ and $y$.
 Pr\"ustel and Mack \cite{1:generalGauge:1} proposed to abandon  unitarity as a 
general
 requirement. In this way, the tetrads of general relativity can be interpreted as parts of  gauge fields, and Higgs fields can be interpreted as parts of 
PT in extra directions. The proposal was motivated by a unified theory of complex systems \cite{2:mack:cmp} and by the fact that in discrete calculus and geometry \cite{3:Mueller-Hoissen}, unitarity is not a natural requirement. Starting from an ordinary gauge theory with unitary PT, real space renormalization group (R.G.) flow may require the introduction of nonunitary PT's $\Phi$ because the attempt to integrate out their selfadjoint factors (which multiply  unitary factors in 
their polar decomposition) may lead to nonlocality. 
This mechanism matches with ideas expressed in the context of deconstruction
\cite{3prime:deconstruction}.
In theories with nonunitary PT there is a holonomy group $H$ which is larger than the unitary gauge group $G\subset H$, and $\Phi\in H$. Only $G$ is a local symmetry, but matter fields must form representation spaces of $H$. 

In our model, the irreducible representations of $H$ remain irreducible when restricted to $G$. 

A 1-dimensional model with computable RG-flow was studied by two of us and illustrates the principles \cite{4:LM:1}. When the PT $\Phi $ becomes nonunitary, there appears a Higgs potential $V(\Phi)$.
Under local gauge transformations $g(\cdot)\in G$, PT 
$\Phi=\phi(x,y)\mapsto g(x)\phi(x,y)g(y)^{-1}$. Gauge invariance implies therefore that $V$ is $G$-biinvariant in the sense that
\be V(u_1\Phi u_2)= V(\Phi) \label{biinv} \ee
for all $u_1,u_2\in G$. 

Here we propose a model which exploits these features to explain the origin of the splitting of the masses of quarks of different flavors.
The consideration of some aspects of leptons is left to a separate study, as are GUTs and questions of supersymmetry.

 Quarks of the same weak hypercharge but belonging to different generations are in the same irreducible representation space of $H$, and their masses are determined by minima of $V(\Phi)$. If the theory is an ordinary gauge theory to begin with and the nonunitary PT and their potential $V$ appear only as a result of a RG-flow, then the quark mass ratios are computable in principle (as functions of gauge coupling constants). $V$ possesses a discrete symmetry, and quark masses split if this symmetry is spontaneously broken.

 The model lives on a 5-dimensional space time, called bulk, with two four 
dimensional boundaries $R$ and $L$. 
Spaces of this kind may appear as a result of orbifolding \cite{5:Kawamura}, and the boundaries may be branes of some kind, but we need not commit ourselves on that. 

Right handed 
quarks $q_R^U=(u_R,c_R,t_R)$ and $q_R^D=(d_R,s_R,b_R)$ 
and all right handed leptons live on $R$ and lefthanded quarks $q_L=(d_L,u_L;s_L,c_L;b_L,t_L)$ and left handed leptons live on $L$. There may be Dirac fermions in the bulk, presumably equally many as there are chiral fermions on the boundaries.

The assignment of fermions to boundaries is compatible with the action of the Pati-Salam subgroup $SU(4)_c\times SU(2)_L\times SU(2)_R$ of the GUT group $SO(10)$, but not with the action of $SO(10)$ itself. One may speculate that at scales shorter than the GUT scale, $SO(10)$ is an unbroken local symmetry of the bulk and light fermions migrate to the boundary when $SU(2)_L\times SU(2)_R$ breaks in the bulk, surviving temporarily on the boundaries. Our model is supposed to be valid at still larger scales, where $SU(2)_R$ is also broken on $R$. Migration of modes from bulk to branes has been discussed in the literature \cite{6:Nilles}.

The unitary gauge group $G_R$ of $R$ is $U(1)\times SU(3)_c$ (weak hypercharge and colour).   It is shared by the bulk and by $L$. 
Since the photon interacts equally with righthanded and lefthanded quarks and mediates interactions between them, we need an abelian gauge field which propagates in the bulk that separates them.
In the following we will disregard colour (except for a comment at the end on the possibility of admitting nonunitary trans-bulk colour PT).  The  elements $u_R=e^{i\vartheta/6}$ of $G_R$ are then parametrized by an angle $\vartheta$ and act on
 right handed quarks according to 
\ba
\left(\begin{array}{c} q^D_R\\q^U_R\end{array}\right)
 &\mapsto &
t_R(u_R)\left(\begin{array}{c} q^D_R\\q^U_R\end{array}\right), \nn \\ 
t_R(u_R) &=&
\left(\begin{array}{cc} u_R^D & 0\\0 & u_R^U\end{array}\right)
=
\left(\begin{array}{cc} e^{-i\vartheta /3} & 0\\0 & e^{2i\vartheta/3}\end{array}\right) \label{uR}
\ea
and on left handed quarks as $q_L\mapsto t_L(u_R)q_L$, $t_L(u_R)=e^{i\vartheta/6}$, as in the standard model. There will be no massless modes other than photons as we shall see. 

The unitary gauge group of $L$ is $G_R\times G_L$, with $G_L=SU(2)$. $G_L$ acts on left handed quarks as in the standard model.
 It consists of matrices
\be
u_L = \left(\begin{array}{cc} u_{11}\mathbf{1}& u_{12}\mathbf{1}\\
u_{21}\mathbf{1} & u_{22}\mathbf{1} \end{array}\right)
\nn \ee
where $\mathbf{1}$ is the $3\times 3$ unit matrix.

The unitary gauge group of the bulk is $G_R\times G$, $G=G^D\times G^U=SU(3)\times SU(3)$ or $SO(3)\times SU(3)$
 or $SU(3)\times SO(3)$. We do not know which of the two 
different groups is associated with $U$ or $D$.
The elements of $G$ are unitary $6\times 6 $ matrices of the form
\be 
\phi = \left( \begin{array}{cc}
\phi^D & 0\\ 0 &  \phi^U \end{array} \right)\ .  \label{2b}
\ee 
At the level of effective theory, the PT $\Phi $ in the bulk are nonunitary, and are elements of the bulk-holonomy group $G_R\times H$, with  $H=\mathbf{R_\ast}H^D\times \mathbf{R_\ast}H^U$, where 
$H^D = SL(3,\mathbf{R})$ or $SL(3,\mathbf{C})$, $H^U = SL(3, \mathbf{C}) $, or the other way round.
$\mathbf{R_\ast}H^D$ consists of positive-real multiples of matrices in $H^D$.
 PT's $\Phi$
have the form $\Phi = diag(\rho^D\phi^D u_R^D, \rho^U\phi^Uu_R^U)$  with $\phi^D\in H^D$ etc. , $u_R^{D,U}$ as in eq.(\ref{uR}), and $\rho^D, \rho^U \in \mathbf{R}$.  

If $H$ has two factors, there will be two additional gauge coupling constants $g_D$, $g_U$.

 At least one of the two groups $G^D, G^U$ needs to be a group of complex matrices. Otherwise there will be no $CP$-violation. 
The choice of two different groups would make it understandable why the mass splittings in the $D$-family and in the $U$-family are different. 

The salient feature of the model is that the elements of $G_L$ and of $G$ can both act on $q_L$ but do not commute. As a result, the local $G$-invariance is broken by the $L$-boundary. This symmetry breaking is responsible for the appearance of a Cabbibo-Kobayashi-Maskawa (CKM) matrix. It would appear that also the $G_L$-invariance on $L$ is broken by interaction with the bulk. But it turns out that this is just the customary Higgs mechanism. How the standard model's Higgs doublet emerges when one passes to a $G_L$-covariant description will be seen below.

\begin{figure}[h]
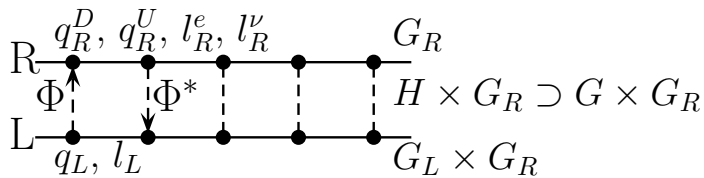

\psset{xunit=0.5cm}
\psset{yunit=0.5cm}
\pspicture(2.5,-2.5)(3,1.5)
\pscircle[fillstyle=solid,fillcolor=black](-4,0.0){0.1}
\pscircle[fillstyle=solid,fillcolor=black](-2,0.0){0.1}
\pscircle[fillstyle=solid,fillcolor=black](0,0.0){0.1}
\pscircle[fillstyle=solid,fillcolor=black](2,0.0){0.1}
\pscircle[fillstyle=solid,fillcolor=black](2,-2){0.1}
\pscircle[fillstyle=solid,fillcolor=black](4,0.0){0.1}
\pscircle[fillstyle=solid,fillcolor=black](-4,-2){0.1}
\pscircle[fillstyle=solid,fillcolor=black](-2,-2){0.1}
\pscircle[fillstyle=solid,fillcolor=black](0,-2){0.1}
\pscircle[fillstyle=solid,fillcolor=black](4,-2){0.1}
\psline[linewidth=.9pt,arrowsize=.20]{-}(-5,0.0)(5,0.0)
\psline[linewidth=.9pt,arrowsize=.20]{-}(-5,-2)(5,-2)
\psline[linewidth=.9pt,linestyle=dashed,arrowsize=.20]{<-}(-4,-0.1)(-4,-1.9)   
\psline[linewidth=.9pt,linestyle=dashed,arrowsize=.20]{-}(0,-0.1)(0,-1.9)
\psline[linewidth=.9pt,linestyle=dashed,arrowsize=.20]{-}(2,-0.1)(2,-1.9)  
\psline[linewidth=.9pt,linestyle=dashed,arrowsize=.20]{-}(4,-0.1)(4,-1.9)  
\psline[linewidth=.9pt,linestyle=dashed,arrowsize=.20]{->}(-2,-0.1)(-2,-1.9)      
\rput[bl](-5.7,-0.3){{\text{\Large R}}}
\rput[bl](-5.7,-2.3){{\text{\Large L}}}
\rput[bl](-4.5,0.3){{\large{$q_R^D$, $q_R^U$, $l_R^e$, $l_R^\nu$}}}
\rput[bl](-4.5,-3){{\large{$q_L$, $l_L$}}}
\rput[bl](4.5,0.3){{\large{$G_R$}}}
\rput[bl](4.5,-3){{\large{$G_L \times G_R$}}}
\rput[bl](-5.0,-1.3){{\Large{$\Phi$}}}
\rput[bl](-1.9,-1.3){{\Large{$\Phi^*$}}}
\rput[bl](4.5,-1.3){{\large{$H \times G_R\supset G \times G_R$}}}
\vspace*{1.3cm}
\endpspicture
\caption{Effective Theory. Assignment of fermions, unitary gauge- and holonomy groups. Nonunitary parallel transporters $\Phi$ transport across the bulk along the dashed
links.}
\label{figure:1}
\end{figure}

Let us now imagine that real space renormalization group transformations have been applied to the theory in the bulk until a lattice spacing in the fifth direction has been reached which equals the width $d$.
Let us assume that $d$ is so small that the gauge theories on the boundaries
can be discretized to lattice gauge theories with lattice spacing $d$ and with appropriate coupling constants without further ado, using appropriate - e.g. domain wall - lattice Dirac-Weyl operators. The resulting effective theory will describe the physics at distances $>>d$.
Except for the $G$-symmetry breaking by the $L$-boundary, it is a defect model of the type examined in ref. \cite{7:Faddeev,8:BAngermann}. It was shown in ref.\cite{8:BAngermann} that the plaquette term is a kinetic term for the Higgs and possesses a natural differential geometric meaning.
The effective theory lives on a lattice with links connecting sites $(\mathbf{x},R)$ on $R$ and $(\mathbf{x},L)$ on $L$, as shown in figure \ref{figure:1}. We write $\mathbf{x}$ in place of $(\mathbf{x},R)$ or $(\mathbf{x},L)$ when it is clear from the context which site is meant. We denote by $\Phi(\mathbf{x})$ the parallel transporter across the bulk along the aforementioned link.    

 Since the RG-transformations must preserve locality, the RG-calculation can be performed to a good approximation in a 5-dimenional theory with $\infty$ extension in the 5-th direction. Therefore the emerging Higgs potential $V(\Phi)$ for $\Phi = diag(\rho^D \phi^D, \rho^U\phi^U)t_R(u_R)\in H\times G_R$, will be $G$-invariant and independent of the factor $t_R(u_R)\in G_R$. Therefore it has the $G$-biinvariance property (\ref{biinv}).

Apart from standard lattice gauge theory terms on the boundaries there will be the following terms in the effective action $S$,
\ba 
S_\Phi & =& \sum_\mathbf{x} 
\left\{\overline{q}_R(\mathbf{x})\Phi(\mathbf{x})q_L(\mathbf{x})\right. \label{Sbulk} \\
&& - \sum_\mu tr \ g^{-2}\left(P_\mu  - 
 \Phi(\mathbf{x})^\ast \Phi(\mathbf{x})\right)  
+ V(\Phi)\} + h.c. \nn \\
P_\mu&=&\Phi(\mathbf{x})^\ast t_R(u_{R\mu}(\mathbf{x}))\Phi(\mathbf{x}+\hat{\mu})t_L(u_{R\mu}(\mathbf{x},L))^\ast u_{L\mu}(\mathbf{x})^\ast \nn
\ea 
plus lepton mass terms,
where $\hat{\mu} $ is the lattice vector in $\mu$-direction on $R$ or $L$, and $g^{-2}=diag(g_D^{-2}, g_U^{-2})$. The
 expression involves PT $u_{R\mu}$, $u_{L\mu}$ in $\mu$-direction
on $R$ and $L$ besides PT $\Phi$. The $P_\mu$-term will be called the plaquette term for short ($\mu = 0...3$). 

It  turns out that the plaquette term is $G$-invariant. The only terms in the action which are not $G$-invariant live on $L$, viz.
$ \sum \overline{q}_L(\mathbf{x}+ \hat{\mu})\gamma^\mu u_{L\mu}(\mathbf{x})^\ast q_L(\mathbf{x}) $,
plus a similar term for leptons.
Our groups are such that elements of $H^D$ can be factored as
\be \phi^D=A^{D\ast}_Rd(\mathbf{\eta}^D)A^D_L \nn \ee
with $A^D_R, A^D_L\in G^D$, and similarly for $\phi^U$. Here $\mathbf{\eta}
=(\eta_1,\eta_2, \eta_3)\in \mathfrak{t}$ lives on the plane $\sum_i \eta_i=0$, $\eta_i$ real,
and 
\be d(\mathbf{\eta})= diag(e^{\eta_1}, e^{\eta_2}, e^{\eta_3}). \nn \ee
Because of $G$-biinvariance, $V$ does not depend on the $A$-factors, whence
\be
V(\Phi)=\mathcal{V}(\rho^D,\rho^U;\mathbf{\eta}^D,\mathbf{\eta}^U). 
\ee 
 Diagonal matrices $d(\eta)$ make up the completely noncompact Cartan subgroup $T^D$ of $H^D$ and $T^U$of $H^U$. To every permutation
$\pi$ of $\{1,2,3\}$ there exists $w^D_\pi\in G^D$ such that 
\be
w_\pi d(\mathbf{\eta})w_\pi^\ast = d(\pi\mathbf{\eta}),
\ee  
where $\pi (\eta_1,\eta_2,\eta_3)=(\eta_{\pi 1},\eta_{\pi 2}, \eta_{\pi_3})$, 
and similarly for $T^U$. It follows from $G$-biinvariance that 
\be \mathcal{V}(\rho^D,\rho^U; \pi \mathbf{\eta}^D, \pi^\prime \mathbf{\eta}^U)
=\mathcal{V}(\rho^D, \rho^U; \mathbf{\eta}^D, \mathbf{\eta}^U).
\label{V_inv} \ee
Pairs $(w_\pi, w_{\pi^\prime})$ form the Weyl group $W_{\mathfrak{t}}$ of the Lie algebra $\mathfrak{t}$ of $T=T^D\times T^U$; it is a discrete group of symmetries of $\mathcal{V}$. The symmetry comes from $G$-invariance. Minima of $\mathcal{V}$ form orbits under $W_{\mathfrak{t}}$. These minima will determine the quark masses. The symmetry is spontaneously broken  if the orbit contains more than one point. If it is completely broken then the quark masses $\{m_d,m_s,m_b\}$ are all distinct, and so are $\{ m_u, m_c, m_t\}$.  This is the situation in the real world. If $(\rho^D_0,\rho^U_0;\mathbf{\eta}^D, \mathbf{\eta}^U)$ is the minimum with $\eta_1^D<\eta_2^D<\eta_3^D$ etc., then 
\ba
 m_d&=&\rho^D_0 e^{\eta^D_1},\  m_s=\rho^D_0 e^{\eta^D_2},\ m_b=\rho^D_0e^{\eta^D_3} , \nn \\
 m_u &=&\rho^U_0 e^{\eta^U_1},\  m_c=\rho^U_0 e^{\eta^U_2},\ m_t=\rho^U_0e^{\eta^U_3} . \nn
 \ea
From the action $S$ of eq.(\ref{Sbulk}), we recover a discretized version of the standard model under the assumption that fluctuations of $\Phi(x)$ away from a $\mathbf{x}$-independent value may be ignored except for  fluctuations of $\rho^U/\rho^U_0$ and $\rho^D/\rho^D_0$ around $1$. In particular, the first term in the action eq.(\ref{Sbulk}) produces quark mass terms.  The factors $A_R^{D,U}$ can be transformed away. Writing $q_L=(q_L^D, q_L^U)^T$, 
\ba \overline{q}_R(\mathbf{x})\Phi(x)q_L(\mathbf{x})&=&
  \overline{q}_R(\mathbf{x})m_D q_L^{D\prime} (\mathbf{x})\rho^D(\mathbf{x})/\rho^D_0 \nn \\
&&
+ \overline{q}_R(\mathbf{x})m_U q_L^{U\prime} (\mathbf{x})\rho^U(\mathbf{x})/ \rho^U_0 \nn
\ea
with $q_{L}^{D \prime}= A^D_L q_{L}^D$, $q_{L}^{U \prime}= A^U_L q_{L}^U$, and mass matrices $m_D=diag(m_d,m_s,m_b)$ etc.. 

The masses are determined by the minima of $V(\Phi)$. 
We will see later how $A_L^{D,U}$ acquire definite values modulo action of the unbroken symmetry $G_{diag}=G^D\cap G^U$. The lattice gauge field $u_R(\mathbf{x})\in U(1)$ attached to the links across the bulk can be gauged away, and the minimization of the $u_R$-factor in the plaquette term constrains
\be
u_R(\mathbf{x},R,\mu) = u_R(\mathbf{x},L,\mu) .
\ee 
There will be fluctuations away from this equality. They are described by a neutral massive vector boson $Z^\prime$ with lattice field
 $w_\mu(\mathbf{x})=u_R(\mathbf{x},R,\mu)
u_R(\mathbf{x},L,\mu)^\ast \ . $
 It is massive because a real multiple of
 the trans-bulk PT $u_R(\mathbf{x})$ acts as a Higgs field 
for it. 

After these fixations, the effective action reduces to a 4-dimensional effective action which is a lattice approximation of the standard model action in unitary gauge, {\em except} that the gauge group is $SU(2)_L\times U(1)_L\times U(1)_R$, which is broken by a Higgs doublet $\varphi$ and a complex scalar Higgs singlet $\xi$. Accordingly there is a neutral massive vector meson  $Z^\prime$ besides $Z$, and a second Higgs particle (if it is not too heavy). Write 
\ba \rho^D(\mathbf{x}) &=&\sigma(\mathbf{x}) \rho(\mathbf{x})\ , \nn \\
\rho^U(\mathbf{x}) &=&\sigma^2(\mathbf{x}) \rho(\mathbf{x})\ .
 \nn 
\ea
 $\rho $ and $\sigma $ are what remains of $\varphi $ and $\xi $ in unitary gauge.

The gauge fixing can be undone as follows. Introduce an 
interface $g(\mathbf{x})\in SU(2)$ between bulk and 
$L$-boundary, fixing it at $1$ initially, and replacing 
$\Phi(\mathbf{x})$ by $\Phi(\mathbf{x})g(\mathbf{x})$. The action is now invariant under $SU(2)$-gauge transformations $v(\mathbf{x})$ which act on $q_L$ and $u_{L\mu}$ as usual, and which take $g(\mathbf{x})\mapsto g(\mathbf{x})v(\mathbf{x})^\ast$. Because of $SU(2)$-gauge-invariance, we may integrate over $g(\mathbf{x})$. 
Given $g(\mathbf{x})$ and $\rho(\mathbf{x})$ define the Higgs doublet field 
$\varphi(\mathbf{x})=\rho (\mathbf{x})\hat{\varphi} (\mathbf{x}) $ as follows.
 To every $g\in SU(2)$ there is a complex 2-vector $\hat{\varphi}$ of unit length such that $g=L[\hat{\varphi}]$ where $L[\hat{\varphi}]\in SU(2)$ is uniquely determined by the requirement that $\hat{\varphi}=L[\hat{\varphi}](0,1)^T.$ It obeys $L[v\hat{\varphi}]= vL[\hat{\varphi}]$ for $v\in SU(2)$. Integrating over $\varphi(\mathbf{x})$ is equivalent to integrating over $g(\mathbf{x})$ and $\rho(\mathbf{x})$. The extra complex 
Higgs field is $\xi(\mathbf{x})=\sigma(\mathbf{x})u_R^D(\mathbf{x}),$
whence $\rho^D\phi^D u^D_R = \xi\phi^D\rho$ and   $\rho^U\phi^U u^U_R = \xi^{\ast 2}\phi^U\rho$.

 The standard models Higgs  potential $V_H$ is given by 
$$
V_H(\rho , \sigma) = min\
\mathcal{V}(\rho\sigma, \rho\sigma^2;
 \mathbf{\eta}^D,\mathbf{\eta}^U).
$$
It depends on two arguments because we have two Higgs fields. 
By definition, the minimum of $V_H$ is at $\rho=\rho_0$, $\sigma=\sigma_0$, where $\sigma_0$ is the vacuum expectation value of $\xi$. But beware: Conventional Higgs fields differ from our $\varphi$, $\xi$ by normalization factors. 

In the presence of leptons one needs extra nonunitary PT's. To give mass to the charged leptons, one needs a PT $\phi^l(\mathbf{x})\in SL(3,\mathbf{C})$ which enters into the charged leptons mass term
 $\propto \overline{e}_R\phi^l l_L$. The most economical choice is 
\be
\phi^l = \phi^D . \label{eco:lepton}
\ee 
This leads to the phenomenologically acceptable mass relation 
$$  m_e:m_\mu:m_\tau = m_d:m_s:m_b \ . $$
To give mass to neutrinos one needs still another PT $\phi^\nu$ and possibly a Majorana mass term for right handed neutrinos, in order to invoke the seesaw mechanism. We refrain from speculating what $\phi^\nu$ might be. 

We add a comment on colour \cite{7:Faddeev}.
Because $G_R$ gauge transformations on $R$ and $L$ are independent,
 there is a colour group $SU(3)_{cL}\times SU(3)_{cR}$ to begin with which is broken to the diagonal $SU(3)_c$ because the $SU(3)_c$ factor in the cross-bulk PT acts as a Higgs field for it. As a result there will be an axigluon \cite{9:axigluon}. 

Let us consider the possibility of admitting nonunitary cross-bulk colour PT $\chi(\mathbf{x})$ in a noncompact colour holonomy group $H_c$ which substitutes for the factor $SU(3)_c$ multiplying $H$. $\chi$ enters as a factor in $\Phi$. Therefore $V(\Phi )$ depends on it. It is a complex $3\times 3$ matrix and admits a 
decomposition 
$\chi =ru_1d(\mathbf{\eta}_c)u_2$ with $u_1$, $u_2\in SU(3)_c$;
let us assume that the factor $r$ is real. Under a gauge transformation $(v_1, v_2)\in SU(3)_c\times SU(3)_c$,
 $\chi \mapsto v_1\chi v_2^\ast$. 
 If the symmetry is to be broken 
down to the diagonal subgroup, the minimum of $V$, considered as a function of $\chi$, must be at $\eta_c=0$, i.e.
 $d(\mathbf{\eta}_c)=\mathbf{1}$ and 
$\chi \in \mathbf{R}_\ast SU(3)_c$. 
The expectation value of the factor $r$, which occurs in the hadronic PT $\phi^D$ but not in its leptonic brother $\phi^l$,
 will determine the ratio between charged lepton masses
 and D-quark masses if the economical choice  
(\ref{eco:lepton}) of the leptonic PT is adopted.

In conclusion, consider $V$ as a function
 of $\mathbf{\eta}$ in  the noncompact Cartan subalgebra $\mathfrak{t}$ of $Lie H$.  The difference between colour and flavor is that the orbits of the minima of $V$ under the Weyl group $W_{\mathfrak{t}}$ of $\mathfrak{t}$ are trivial for colour and nontrivial for flavor. In other words the $W_{\mathfrak{t}}$-symmetry is spontaneously broken for flavor, but not for colour.

Let us finally turn to the CKM matrix.
 Assuming there are really two independent factors $G^D$, $G^U$ in $G$, the CKM matrix $C$ could be transformed away if it were not for the breaking of $G$-invariance by the $L$-boundary. $C$ is therefore not determined by the minima of $V$ but could be obtained as follows.

$A^D_L(\mathbf{x})$ and $A^U_L(\mathbf{x})$ are dynamical fields of the effective theory, because $\Phi(\mathbf{x})$ depends on them. Let $W(A^D_L,A^U_L)$ be the effective action for these fields alone obtained by 
integrating out the quark-fields and gauge fields associated with $G_R$ and
 $G_L$, as well as $A^D_R, A^U_R$ and the fluctuations of $\mathbf{\eta}^D, \mathbf{\eta}^U$ and of $\rho^D, \rho^U$ away from the minimum of $V$.  The quark masses determined by minimization of $V$ enter as parameters into $W$. The CKM field 
$$ C(\mathbf{x})= A^D_L(\mathbf{x})A^{U\ast}_L(\mathbf{x}) $$
is invariant under global $G_{diag}$-transformations, and so is $W$. The CKM matrix $C$ is determined by the ground state of the theory with action $W$. In tree approximation (which may be accurate enough or not), it is determined by the minimum of the restriction of $W$ to constant fields $C$; this is a calculation essentially within the standard model.  It would be interesting to deal with 
the full effective action by numerical means.

One may consider more complicated models where higher dimensional spaces with boundaries intersect such that the boundaries' intersections are 4-dimensional. 
String theorists are invited to exhibit suitable branes and predict how much supersymmetry there should be in the bulk and on the boundaries. 

It is a pleasure to thank B. Angermann, M. Olschewski, F. Neugebohrn,  M. de Riese and M. R\"ohrs for discussions, and Laura Covi for asking the right questions. C.L. and T.P. thank Deutsche Forschungsgemeinschaft for financial support.\\[4mm]

\end{document}